\documentclass[10pt,aps,prd,showpacs,nofootinbib,floats,floatfix,preprintnumbers,groupedaddress,twocolumn]{revtex4-2}

\usepackage{aas_macros}
\usepackage{cancel}

\usepackage[normalem]{ulem} 
\usepackage{bm}
\usepackage[titletoc]{appendix}
\usepackage{hyperref}
\usepackage{latexsym}
\usepackage{dcolumn}
\usepackage{amsmath,amsfonts,amssymb}
\usepackage{graphicx,epsfig}
\usepackage{amsmath}
\usepackage{fancyhdr}
\usepackage{hyperref}
\usepackage{graphicx,epstopdf}
\usepackage{xcolor}
\usepackage{siunitx}

\begin{document}

\title{Neutrino-dominated relativistic shocked accretion flow around rotating black hole: implications for short gamma-ray bursts}

\author{Amit Kumar}\email{kamit@iitg.ac.in}
\author{Sayan Chakrabarti}\email{sayan.chakrabarti@iitg.ac.in}
\author{Santabrata Das}\email{sbdas@iitg.ac.in (Corresponding Author)}

\affiliation{Indian Institute of Technology Guwahati, Guwahati, 781039, Assam, India}
\date{\today}

\begin{abstract}

We investigate the physical properties of the central engine powering gamma-ray bursts (GRBs), modelled as a stellar-mass black hole accreting via a neutrino-dominated accretion flow (NDAF). By solving the governing hydrodynamic equations, we obtain global transonic NDAF solutions featuring shock transitions and examine their role in powering GRB energetics. The NDAF solutions are explored over a broad range of black hole parameters, including its mass ($M_{\rm BH}$) and spin ($a_{\rm k}$), and accretion rate ($\dot{M}$). We find that shocked NDAFs can naturally account for the observed diversity in GRB energy output. Incorporating results from numerical simulations of binary neutron star and black hole–neutron star mergers, we estimate the remnant black hole mass and spin parameters for the predicted range of post-merger disk mass ($M_{\rm disk}$). Our analysis reveals that small-mass black holes with relatively low spin values can adequately reproduce the luminosities of short GRBs (SGRBs), whereas identical GRB luminosities can also be achieved for more massive black holes possessing higher spin values. Finally, we uncover a robust correlation between the black hole spin and disk mass such that $M_{\rm disk}$ decreases with increasing $a_{\rm k}$, remaining largely independent of the black hole mass ($M_{\rm BH}$) powering GRBs.

\end{abstract}

\maketitle


\section{\label{Introduction}Introduction}

Gamma-ray bursts (GRBs) are among the most luminous and transient phenomena in the universe typically lasting only a few seconds. Based on the observed burst duration, GRBs are broadly categorized into short gamma-ray bursts (SGRBs; $T_{90} < 2 {\rm s}$) and long gamma-ray bursts (LGRBs; $T_{90} > 2 {\rm s}$) \cite[]{Kouvelioto-1993}, where $T_{90}$ denotes the time interval during which the burst fluence rises from $5\%$ to $95\%$ above the background level \cite[]{Zhang-Meszaros-2004, Nakar-2007}. It is widely believed that the central engine of GRBs is powered by a stellar-mass black hole surrounded by a hyperaccretion disk with an accretion rate of $0.001$–$10~M_{\odot}~ {\rm s}^{-1}$ \cite[]{Paczynski-1991, Narayan-etal-1992}. Under such extreme conditions, the accreting plasma becomes exceedingly hot and dense ($\rho \sim 10^8–10^{12}~{\rm g~cm}^{-3}$, $T \sim 10^{10}–10^{11}~{\rm K}$), resulting in a high optical depth photon-trapping environment where photons cannot efficiently escape, rendering them ineffective in powering the observed emission. However, such conditions are conducive to the efficient production of neutrinos and antineutrinos, which carry away the accretion energy from the accretion disk. After escaping the disk, these neutrinos and antineutrinos subsequently annihilate above the disk to generate an electron–positron pair-dominated outflow capable of powering GRBs. Such neutrino-dominated accretion flows (NDAFs) have been extensively studied in the literature \cite[]{Popham-etal-1999, Narayan-etal-2001, DiMatteo-etal-2002, Kohri-minesgige-2002, Kohri-etal-2005, Gu-etal-2006, Chen-Beloborodov-2007, Liu-etal-2007, Kawanaka-Mineshige-2007, Xue-etal-2013, Cao-etal-2014, Xie-etal-2016, Song-etal-2020, Chen-etal-2022, She-etal-2022, Kumar-etal-2025}. In these studies, the structure of NDAFs is computed by incorporating neutrino emission processes, and the GRB energetics are estimated via neutrino annihilation. An alternative mechanism to power the GRB involves extracting the spin energy of the central black hole and converting it into the kinetic energy of the outflow \cite[]{Blandford-Znajek-1977}.

Energetic events such as merger of binary neutron stars (NS+NS) \cite[]{Paczynski-1986, Eichler-etal-1989}, neutron star and black hole (NS+BH) \cite[]{Paczynski-1991}, and collapsars \cite[]{Woosley-1993, MacFadyen-Woosley-1999} are considered as the potential progenitors of GRBs. These events generally result in the formation of a stellar-mass black hole surrounded by a hyperaccreting disk. Observation of kilonova emissions \cite[]{Tanvir-etal-2013, Troja-etal-2019, Jin-etal-2020} produced by the decay of radioactive species formed during the merger process and gravitational wave signals generated by inspiraling compact objects provides strong evidence linking SGRBs to neutron star mergers \cite[]{Abbott-etal-2017}. In contrast, LGRBs are typically associated with supernovae \cite[]{Galama-etal-1998, Stanek-etal-2003, Kelly-etal-2008, Svensson-etal-2010} and are most often found in star-forming regions \cite[]{Totani-1997, Bloom-etal-1998}. Following the merger, the accretion of matter from the remnant disk fuels relativistic outflows through previously discussed mechanisms ultimately producing the gamma-ray burst. After the prompt burst, the ejecta interact with the surrounding medium and lose energy producing longer-wavelength emission known as the “afterglow”, which can be observed over weeks to months \cite[]{Costa-etal-1997, vanParadijs-etal-1997, Bremer-etal-1998, Heng-etal-2008}. However, the central black hole remains obscured by intense radiation, making it difficult to constrain key physical parameters such as its mass ($M_{\rm BH}$), spin ($a_{\rm k}$), and mass accretion rate ($\dot{M}$) through direct observation.

In parallel with observational efforts, numerical simulations play a crucial role in advancing the understanding of compact object mergers and their post-merger evolution. A wide array of studies has investigated how the initial parameters, namely the mass ratio of the binary system, the neutron star equation of state, spin, and magnetic field strength, influence the dynamical outcome of the merger. These simulations provide valuable insights into the physical properties of the remnant system and can help in constraining the characteristics of the final central black hole. Simulation results predominantly indicate that the merger remnant eventually collapses into a stellar-mass black hole surrounded by an accretion disk with disk masses ranging from $\sim 0.01$ to several tenths of a solar mass \citep{Kluzniak-Lee-1998, Ruffert-Janka-1998, Lee-Kluzniak-1999, Janka-etal-1999, Ruffert-Janka-2001, Shibata-etal-2003, Shibata-etal-2005, Shibata-Taniguchi-2006, Oechslin-Janka-2006, Kiuchi-etal-2009, Rezzolla-etal-2010, Hotokezaka-etal-2013, East-etal-2016, Most-etal-2019, Ruiz-etal-2020, Tootle-etal-2021, Sun-etal-2022, Cokluk-etal-2024}. This range of disk mass places important constraints on the mass accretion rate onto the black hole, which directly influences the energy output available to power the GRBs.

Given the plausible supply of matter in the aforementioned scenarios, numerous studies have explored the viability of neutrino-dominated accretion flows (NDAFs) as central engines for both short and long GRBs \citep{Fan-Wei-2011, Liu-etal-2015, Song-etal-2016, Xie-etal-2016}. These investigations demonstrate that, through neutrino–antineutrino annihilation, NDAFs are capable of powering the observed GRB energetics. However, \citet{Liu-etal-2015, Song-etal-2016} argued that while a majority of GRBs can be explained within the typical disk mass range, certain bursts may require a more massive accretion disk to match the observed luminosities. It is worth noting that these models generally assume a Keplerian disk structure \citep{Shakura-Sunyaev-1973, Chen-Beloborodov-2007}, which results in a subsonic flow configuration. However, to satisfy the inner boundary condition at the event horizon, the accreting matter must approach the black hole at the speed of light, that demands the NDAF solution to be transonic in nature \cite[]{Kumar-etal-2025}. Depending on the flow parameters, NDAF may undergo a discontinuous transition of the flow variables in the form of a shock, provided the Rankine–Hugoniot conditions are satisfied \cite[]{Landau-Lifshitz-1959}. From the thermodynamic perspective, these shock solutions are favored due to their higher entropy content \cite[]{Kumar-etal-2025}. Moreover, these shock-induced NDAF solutions effectively account for the wide range of energies observed in GRBs. It is noteworthy that low angular momentum, advective NDAF models exhibit a notable degeneracy, where different combinations of mass accretion rate ($\dot{M}$) and black hole spin ($a_{\rm k}$) seems to yield similar GRB luminosities. This inherent degeneracy highlights the need for further theoretical investigation, particularly through the incorporation of shock dynamics within the NDAF framework. When combined with constraints from merger simulations, such an approach provides a promising avenue for reducing the admissible parameter space of the GRB central engine.

Being motivated by this, we revisit the shock formalism in NDAFs and assess its role in powering short GRBs by incorporating results from compact binary merger simulations. We adopt an effective potential that approximates the space-time geometry around a rotating black hole \cite[]{Dihingia-etal-2018} and solve the NDAF governing equations to obtain global transonic solutions with shocks. Using these solutions, we compute the neutrino annihilation luminosity ($L_{\nu \bar{\nu}}$) and compare it with the observed energetics of selected SGRBs. To test the viability of the model, we explore a wide range of black hole masses ($M_{\rm BH}$) and spin parameters ($a_{\rm k}$), and fit the model luminosities to observed values. Finally, by incorporating the expected disk mass from merger simulations, we place constraints on the spin of the central engine powering SGRBs.

The structure of this paper is as follows. In Section~\ref{Gov_eq}, we outline the model assumptions and introduce the governing equations describing neutrino-dominated accretion flows. In Section~\ref{RESULTS}, we present the transonic accretion solutions, including the associated flow variables, and discuss their implications for a sample of short GRBs. Finally, in Section~\ref{discussion_summery}, we summarize our results.

\section{\label{Gov_eq}Governing equations}

In the quest to unravel the dynamics of hyperaccretion flows powering short gamma-ray bursts (SGRBs), we consider a steady, axisymmetric, viscous, advective accretion flow around a rotating black hole. The flow is assumed to be confined to the equatorial plane and it maintains hydrostatic equilibrium along the vertical direction. To account for gravitational effects, we use an effective potential that satisfactorily describes the space-time geometry around the rotating black hole \cite[]{Dihingia-etal-2018}. The analysis is carried out in a cylindrical coordinate system with the black hole at the origin, adopting a unit system where $G=M_{\rm BH}=c=1$, with $G$ denoting the gravitational constant, $M_{\rm BH}$ the black hole mass, and $c$ the speed of light. In this framework, length, angular momentum, and energy are expressed in units of $r_{\rm g}=GM_{\rm BH}/c^2$, $r_{\rm g}c$, and $c^2$, respectively.

The governing equations of hydrodynamics that describe fluid motion around a rotating BH are given by
\noindent (a) Radial momentum equation:
\begin{equation}
v\frac{dv}{dx}+\frac{1}{\rho}\frac{dP}{dx}+\frac{d \Phi_{\rm {eff}}}{dx}=0, 
 \label{rad_mom_eq}
 \end{equation}
 where $x$, $v$, $P$ and $\rho$ denote the radial coordinate, radial velocity, total pressure and mass density of the accretion flow. The term $\Phi_{\rm eff}$ represents the effective potential on equatorial plane \cite[]{Dihingia-etal-2018} and is given by,
\begin{equation}
    \Phi_{\rm {eff}}= \frac{1}{2} \ln\left[\frac{x \Delta}{a^2_{\rm{k}}(x+2)-4 a_{\rm k} \lambda+x^3-\lambda^2(x-2)}\right]\label{eff_potential},
\end{equation}
where $\Delta=x^2-2x+a_{\rm k}^2$, $\lambda$ represent the angular momentum of flow, and $a_{\rm k}$ is the spin of the black hole.
 
\noindent (b) Mass conservation equation:
\begin{equation}
\dot M=4\pi v \rho H \sqrt{\Delta},\label{mass_cons_eq}
\end{equation}
where $\dot{M}$ denotes the mass accretion rate, which remains constant throughout the flow in absence of massloss from the disk, and $H$ is the local half-thickness of the disk. For convenience, in this work, we express the accretion rate in terms of solar mass per second as $\dot{m}= \dot{M}/ M_{\odot}~\rm s^{-1}$. Following \cite[]{ Riffert-Herold-1995, Peitz-Appl-1996}, we obtain $H$ as
\begin{equation}
    H =\sqrt{\frac{P x^3 }{\rho \mathcal{F}}},~ \mathcal{F}=\frac{1}{1-\lambda \Omega} \times \frac{(x^2 +a_{\rm k}^2)^2 + 2 \Delta a_{\rm k}^2}{(x^2 +a_{\rm k}^2)^2 - 2 \Delta a_{\rm k}^2},
\end{equation}
where $\Omega~\left[= (2 a_{\rm k} +\lambda(x-2))/(a_{\rm k}^2(x+2)-2 a_{\rm k}\lambda +x^3)\right]$ is angular velocity of flow. We define the sound speed of the flow as $C_{\rm s}=\sqrt{P/\rho}$, where $P=P_{\rm gas}+P_{\rm rad}$ with $P_{\rm gas} \left(= \rho k_{\rm B} T/m_{\rm p}\right)$ being the gas pressure and $P_{\rm rad} \left( = 11 \bar{a} T^4/12 \right)$ being the pressure due to radiation \cite[]{DiMatteo-etal-2002}. Here, $k_{\rm B}$, $\bar{a}$ and $m_{\rm p}$ are the Boltzmann constant, radiation constant and mass of the proton, respectively.
  
\noindent (c) Azimuthal momentum equation:
\begin{equation}
\frac{d\lambda}{dx}+\frac{1}{ \Sigma v x}\frac{d}{dx}(x^2 W_{x\phi})=0,
\label{azimuthal_eq}
\end{equation}
In equation (\ref{azimuthal_eq}), $W_{x \phi}$ refers $x \phi$ component of viscous stress and is given by \cite[]{Chakrabarti-Molteni-1995}, 
$$W_{x \phi}=-\alpha(W + \Sigma v^2),$$
where $\alpha$ denotes the viscosity parameter, and $W~(= 2 P H)$ and $\Sigma~(=2 \rho H)$ represent the vertically integrated pressure and density, respectively.

\noindent (d) Entropy generation equation: 
\begin{equation}
    \Sigma v T\frac{ds}{dx}=\Sigma v  \left(\frac{du}{dx}-\frac{P}{\rho^2}\frac{d\rho}{dx}\right)=Q^{-}-Q^{+},
    \label{entrop_gen_eq}
\end{equation}
where $T$ and $s$ are the temperature and entropy of the flow, and $u$ represents the specific internal energy which is given by,
\begin{equation}
    u=\frac{P_{\rm gas}}{\rho(\gamma-1)} + \frac{3 P_{\rm rad}}{\rho},
    \label{internal_energy}
\end{equation}
with $\gamma$ being the adiabatic index, which we choose as $4/3$ all throughout unless specified otherwise. It is worth mentioning that in equation \eqref{internal_energy}, we refrain from including electron degeneracy pressure and neutrino pressure. In realistic NDAFs, electrons remain typically mildly or moderately degenerate at relatively high accretion rates (${\dot M} \sim M_\odot~{\rm s}^{-1}$) \cite[]{Kawanaka-Mineshige-2007,Chen-Beloborodov-2007,Liu-etal-2007,Xue-etal-2013}. Furthermore, at such accretion rates, neutrinos escape efficiently rendering their pressure contribution subdominant \cite[]{DiMatteo-etal-2002,Kohri-etal-2005,Xie-etal-2016}. Since the present analysis is restricted to ${\dot M} \le M_\odot~{\rm s}^{-1}$, we argue that equation \eqref{internal_energy} yields reliable estimates of the thermodynamic quantities and neutrino luminosities, ensuring that the overall conclusions regarding the shock-induced NDAF solutions remain unaffected. Moreover, $Q^{+}$ denotes the heat generation by viscous heating and is expressed as \cite[]{Chakrabarti-1996, Chakrabarti-Das-2004},
\begin{equation}
    Q^{+} \equiv Q^{\rm vis}= x W_{x \phi} \frac{d\Omega}{dx}.
\end{equation}
In this work, the cooling of the flow is governed by the neutrinos ($\nu$) \cite[]{DiMatteo-etal-2002}, and we have
\begin{equation}
     Q^{-}= Q_\nu=\sum_{\rm{i}}\frac{(7/8)\sigma T^4}{(3/4)[\tau_{\rm{\nu_{i}}}/2+1/\sqrt{3}+1/(3\tau_{\rm{a,\nu_{i}}})]},
     \label{neutrino_cooling}
\end{equation}
where $\sigma$ is the Stefan-Boltzmann constant, and $\tau_{\rm{\nu_{i}}}~\left( =\tau_{\rm{a,\nu_{i}}} +\tau_{\rm{s,\nu_{i}}}\right)$ is the sum of absorption and scattering optical depth for each neutrino flavor $(\rm{i} \rightarrow e,\mu,\tau)$. We introduce dimensionless temperature ($\Theta~= k_{\rm B} T/m_{\rm e} c^2$), where $m_{\rm e}$ is mass of electron.

We simplify equations (\ref{rad_mom_eq}), (\ref{mass_cons_eq}), (\ref{azimuthal_eq}) and (\ref{entrop_gen_eq}) and obtain the wind equation as 
 
\begin{equation}
    \frac{dv}{dx}=\frac{N(x,v,\lambda,\Theta,a_{\rm k}, \dot{m})}{D(x,v,\lambda,\Theta, a_{\rm k}, \dot{m})}\label{dvdr},
\end{equation}
where the numerator $N$ and denominator $D$ are the explicit functions of the flow variables and are given in the Appendix. We further calculate the gradient of angular momentum ($d\lambda/dx$) and gradient of dimensionless temperature ($d\Theta/dx$) in terms of $dv/dx$, which are given by,
\begin{equation}
\frac{d\lambda}{dx}=\lambda_{11}\frac{dv}{dx}+\lambda_{12},
\label{dldr}
\end{equation}       
and 
\begin{equation}
\frac{d\Theta}{dx}=\Theta_{11}\frac{dv}{dx}+\Theta_{12}.
\label{dthetadr}
\end{equation}     
In equations (\ref{dldr}) and (\ref{dthetadr}), the coefficients $\lambda_{11}$, $\lambda_{12}$, $\Theta_{11}$ and $\Theta_{12}$ are expressed in terms of the flow variables and presented in Appendix.

To obtain the accretion solutions, we solve equations (\ref{dvdr}), (\ref{dldr}), and (\ref{dthetadr}) simultaneously using the 4th order Runge-Kutta method, following the methodology outlined by \cite[]{Chakrabarti-Das-2004, Das-2007}. We use the accretion rate ($\dot{m}$), viscosity parameter ($\alpha$), mass of black hole ($M_{\rm BH}$), and black hole spin ($a_{\rm k}$) as global parameters, while the local flow variables at the critical point ($x_{\rm c}$), such as the energy ($\mathcal{E}_{\rm c}$) and angular momentum ($\lambda_{\rm c}$), serve as local parameters. The local energy is computed using $\mathcal{E}(x) = v^2/2 + h + \Phi_{\rm eff}$, where $h$ is the specific enthalpy of the flow ($h = u + P/\rho$). This approach is essential because the accretion flow around a black hole is inherently transonic and must pass through a critical point before entering the black hole. Consequently, using these global and local parameters, we integrate equations (\ref{dvdr}), (\ref{dldr}), and (\ref{dthetadr}) inward from $x_{\rm c}$ to the horizon, and outward to the outer edge ($x_{\rm edge}$). Upon joining these two parts, we obtain the global transonic accretion solution around black hole. 

Depending on the flow parameters, the accretion flow may exhibit single or multiple critical points \cite[]{Das-etal-2001, Chakrabarti-Das-2004, Das-2007}. The critical point formed close to the horizon is termed the inner critical point ($x_{\rm in}$), whereas the outer critical point ($x_{\rm out}$) is located farther from the black hole horizon. Typically, the inner critical point ($x_{\rm in}$) and the corresponding angular momentum ($\lambda_{\rm in}$) range from $1.5 \lesssim x_{\rm in} \lesssim 3$ and $2 \lesssim \lambda_{\rm in} \lesssim 4$, respectively \cite[]{Das-Chakrabarti-2004}. Thus, we provide the local flow parameters at $x_{\rm in}$ to obtain the complete global accretion solutions. Notably, an accretion solution can pass through both $x_{\rm in}$ and $x_{\rm out}$, provided the flow possesses a shock wave \cite[]{Das-etal-2001, Das-Chakrabarti-2004}.

\section{\label{RESULTS}Result}

\subsection{Global shocked NDAF solutions}

\begin{figure}
    \centering
    \includegraphics[width=\columnwidth]{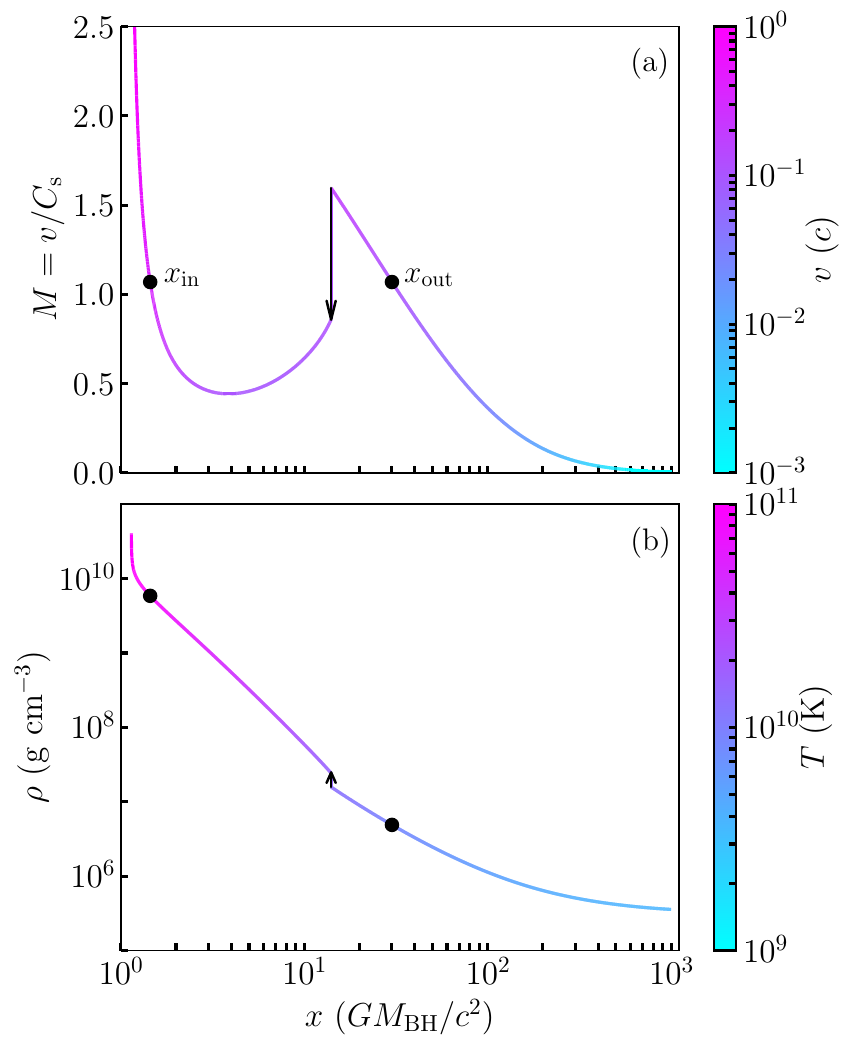}
    \caption{Variation of ($\rm a$) Mach number ($M = v/C_{\rm s}$) and ($\rm b$) density ($\rho$) of the NDAF as functions of radial distance ($x$). The corresponding velocity ($v$) and temperature ($T$) profiles are indicated by the color, with the color bars on the right denoting their respective ranges. The flow parameters are chosen as $\mathcal{E}_{\rm in} = 0.0001$, $\lambda_{\rm in} = 1.9880$, $a_{\rm k} = 0.99$, $\dot{m} = 0.01$, $M_{\rm BH} = 3 M_{\odot}$, and $\alpha = 0.001$, respectively. The vertical arrow at $x_{\rm s} = 14.0258$ indicates the location of the shock transition, while the filled circles indicate the location of the inner ($x_{\rm in}$) and outer ($x_{\rm out}$) critical points. See text for the details.
    }
    \label{fig1}
\end{figure}

In Fig. \ref{fig1}, we show a typical example of the global transonic accretion solution for NDAFs, where radial profiles of different flow variables are depicted. For this solution, the global flow parameters are set to $\dot{m} = 0.01$, $\alpha = 0.001$, $M_{\rm BH} = 3 M_{\odot}$, and $a_{\rm k} = 0.99$. In addition, we choose the local flow parameters at the inner critical point ($x_{\rm in}=1.4468$) as $\mathcal{E}_{\rm in}=0.0001$ and $\lambda_{\rm in}=1.9880$. In panel ($\rm a$), we plot the variation of Mach number ($M=v/C_{\rm s}$) with radial distance ($x$), together with the behaviour of velocity ($v$), denoted with color, as indicated by the color bar. As shown in figure, the flow starts from the outer edge of the disk at $x_{\rm {edge}}=1000$ with a subsonic speed. Due to the influence of the black hole's gravity, the radial velocity of the flow increases and reaches supersonic speeds after passing the outer critical point at $x_{\rm out}=30.0864$. After becoming supersonic, the flow encounters a centrifugal repulsion, causing the matter to accumulate around the black hole. Upon reaching a certain threshold of accumulation, the centrifugal barrier triggers a discontinuous transition in the flow variables, which manifests as the formation of shock waves. For flows with hydrostatic equilibrium, the standing shock forms provided the Rankine–Hugoniot conditions (RHCs) are satisfied \cite[]{Landau-Lifshitz-1959} which are expressed as 
\begin{align}
    &\text{Energy flux conservation}:~\mathcal{E}_{+}=\mathcal{E}_{-},\label{SC_ENERGY_FLUX}\\
    &\text{Mass flux conservation}:~\dot{M}_{+}=\dot{M}_{-},\label{SC_MASS_FLUX}\\
    &\text{Momentum flux conservation}:~\Pi_{+}=\Pi_{-},\label{SC_MOMENTUM_FLUX}
\end{align}
where the subscripts `$-$' and `$+$' represent the flow variables just before and after the shock, respectively, and $\Pi ~(= W + \Sigma v^2)$ denotes the total vertically integrated pressure.
For the above chosen flow parameters, we find the shock transition to occur at the location $x_{\rm s} = 14.0258$, where the RHCs are satisfied. After the shock, the flow continues to accrete and ultimately falls into the black hole after passing through the inner critical point at $x_{\rm in}=1.4468$. In the figure, this shock transition is indicated by the vertical arrow, and the inner ($x_{\rm in}$) and outer ($x_{\rm out}$) critical points are marked by filled circles. For the shocked solution, the matter makes transition from supersonic to subsonic branch at $x_{\rm s}$, where shock compression leads to an increase in density, as illustrated in panel (b). Furthermore, the temperature of the post-shock flow (equivalently post-shock corona, hereafter PSC) increases as the kinetic energy of the pre-shock flow is converted into thermal energy. The overall temperature variation of the flow is shown using color where the color bar on right indicate the temperature range. This rise in density and temperature across the shock front leads to an overall enhancement in entropy at PSC, making the shock-induced accretion solution thermodynamically more favorable compared to solutions without shock \cite[]{Becker-Kazanas-2001}.

Such high post-shock temperature and density conditions naturally facilitate substantial energy losses through the emission of neutrinos and antineutrinos. The subsequent annihilation of these particles can drive relativistic outflows and potentially serve as the energy source for the formation of a baryonic fireball \cite[]{Eichler-etal-1989}. Given the detailed density and temperature profiles of the shock solution, we calculate the annihilation luminosity ($L_{\nu \bar{\nu}}$) using the equation \cite[]{Xue-etal-2013,Kumar-etal-2025}, 
\begin{equation}
    L_{\nu \bar{\nu}} = 4 \pi\int^{\infty}_{x_{i}}\int^{\infty}_{H} l_{\nu\bar{\nu}}\,x \,dx \,dz, \label{Neutrino_anni_lum}
\end{equation}
Here, $x_{i}$ and $z$ represent the inner edge of the disk and the vertical distance from the disk plane, respectively. The term $l_{\nu\bar{\nu}}$ refers the energy deposition rate per unit volume due to neutrino annihilation \cite[]{Ruffert-etal-1997,Popham-etal-1999, Rosswog-etal-2003,Kumar-etal-2025}. Next, we compute the total annihilation luminosity, $L_{\nu \bar{\nu}}$, for a range of model parameters to estimate the energy output from neutrino annihilation in GRBs event. This analysis, in turn, enables us to constrain key physical properties of the central engine, namely its mass and spin, which we discuss in detail in the next section.

\subsection{Astrophysical implications}

We apply the shock-based model formalism within the neutrino dominated accretion flow (NDAF) framework to estimate the spin ($a_{\rm k}$) of the central black hole. In doing so, we compute the mean fireball output power from the central engine ($\dot{E}$) of the GRB \cite[]{Fan-Wei-2011,Liu-etal-2015} as,
\begin{equation} 
\dot{E} \approx \frac{(1 + z^{\prime})(E_{\rm \gamma, iso} + E_{\rm k, iso}) \theta_{\rm j}^{2}}{2 T_{90}},\label{dotE} 
\end{equation}
where $E_{\rm \gamma, iso}$ is the isotropic radiative energy emitted in the $\gamma$-ray band, $E_{\rm k, iso}$ is the isotropic kinetic energy powering the afterglow phase, $z^{\prime}$ is the redshift, $T_{90}$ is the duration of burst, and $\theta_{\rm j}$ is the opening angle of ejecta. Simulation studies suggest that the quantity $\dot{E}$ represents the fraction of total neutrino annihilation luminosity. Following the approach of \cite{Aloy-etal-2005}, we estimate the observed neutrino annihilation luminosity ($L^{\rm obs }_{\nu\bar{\nu}}$) associated with GRBs as,
\begin{equation}
L^{\rm obs }_{\nu\bar{\nu}} = \frac{{\dot E}}{\eta},\label{mean_out_pow}
\end{equation}
where $\eta$ denotes the efficiency with which neutrino-antineutrino annihilation energy is converted into the fireball that powers the GRB. As the exact value of $\eta$ remains uncertain and is not well constrained, we adopt a commonly used representative value of $\eta = 0.1$ for our calculations \cite[]{Liu-etal-2015,Kumar-etal-2025}. We compile observational data for the selected short GRBs from the literature, namely $T_{90}$, $z^{\prime}$, $E_{\rm \gamma, iso}$, $E_{\rm k, iso}$, $\theta_{\rm j}$, and use equation (\ref{mean_out_pow}) to compute the corresponding values of $L^{\rm obs}_{\nu\bar{\nu}}$, as listed in Table \ref{table-1}.

\begin{table*}
\caption{Observation details of the SGRB sources under consideration. Columns $1-8$ denote source name, duration of burst ($T_{90}$), redshift ($z^{\prime}$), isotropic radiative energy ($E_{\rm \gamma,iso}$), isotropic kinetic energy ($E_{\rm k,iso}$), opening angle ($\theta_{\rm j}$), observed neutrino annihilation luminosity ($L^{\rm obs}_{\nu \bar{\nu}}$), and references. See text for the details.}
\begin{ruledtabular}
\begin{tabular}{lccccccl}
Source & $T_{90}$ & $~z^{\prime}$ & $E_{\rm \gamma,iso}$ & $E_{\rm k,iso}$ & $\theta_{\rm j}$ & $ L^{\rm obs}_{\nu \bar{\nu}}$ & References\\
    & (s) & & $(\times 10^{51}~\rm erg)$ & $(\times 10^{51}~\rm erg)$ & ($\rm rad$) & $\rm (erg~s^{-1})$ & \\
\hline
GRB 131001A & 1.54 & 0.7170 & 0.370 &  5.41 & $\sim 0.05$ & $8.06\times10^{49}$ & \cite{Liu-etal-2015} \\
GRB 051221A & 1.40 & 0.5465 & 0.920 & 12.60 & $\sim 0.12$ & $1.07\times10^{51}$ & \cite{Soderberg-etal-2006} \\
GRB 050724  & 3.00 & 0.2570 & 0.100 &  0.27 & $\sim 0.35$ & $9.50\times10^{49}$ & \cite{Grupe-etal-2006} \\
GRB 070714B & 2.00 & 0.9230 & 1.610 &  2.32 & $\sim 0.05$ & $4.72\times10^{49}$ & \cite{Liu-etal-2015} \\
GRB 070809	& 1.30 & 0.4730 & 0.056 &  3.91 & $\sim 0.05$ & $5.62\times10^{49}$ & \cite{Liu-etal-2015} \\
\end{tabular}
\end{ruledtabular}
\label{table-1}
\end{table*}
 
\begin{figure}
    \centering
       \includegraphics[width=\columnwidth]{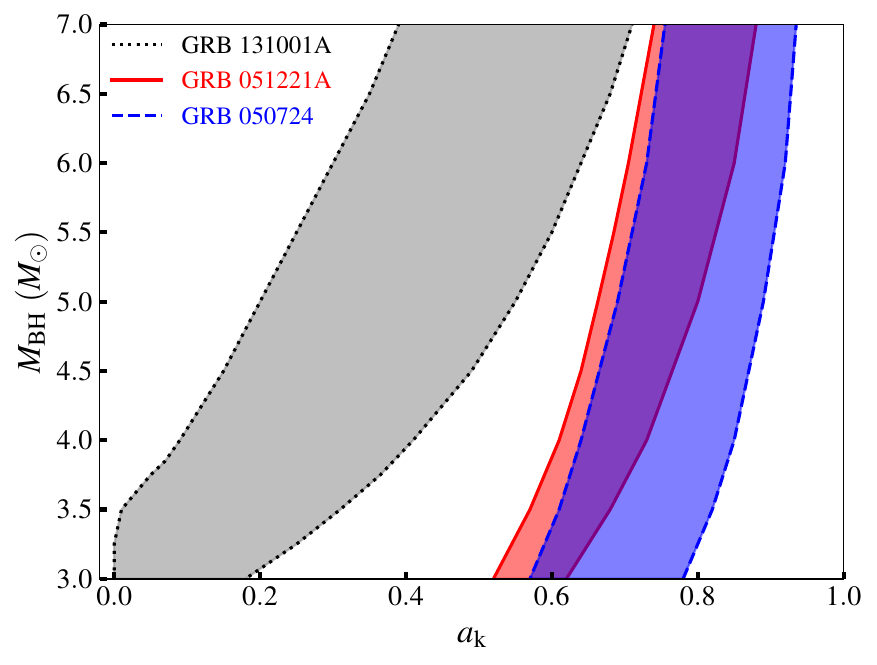}
    \caption{Parameter space in the $a_{\rm k}–M_{\rm BH}$ plane showing the allowed ranges of black hole spin ($a_{\rm k}$) and mass ($M_{\rm BH}$) yielding the observed $L^{\rm obs}_{\nu {\bar \nu}}$. Grey, red, and blue shaded regions denote obtained results for GRB 131001A, GRB 051221A, and GRB 050724, respectively. Here, we choose $M_{\rm disk} = 0.2 M_{\odot}$ and $\alpha = 0.01$. See text for the details.
    }
    \label{fig2}
\end{figure}

\begin{figure}
    \centering
    \includegraphics[width=\columnwidth]{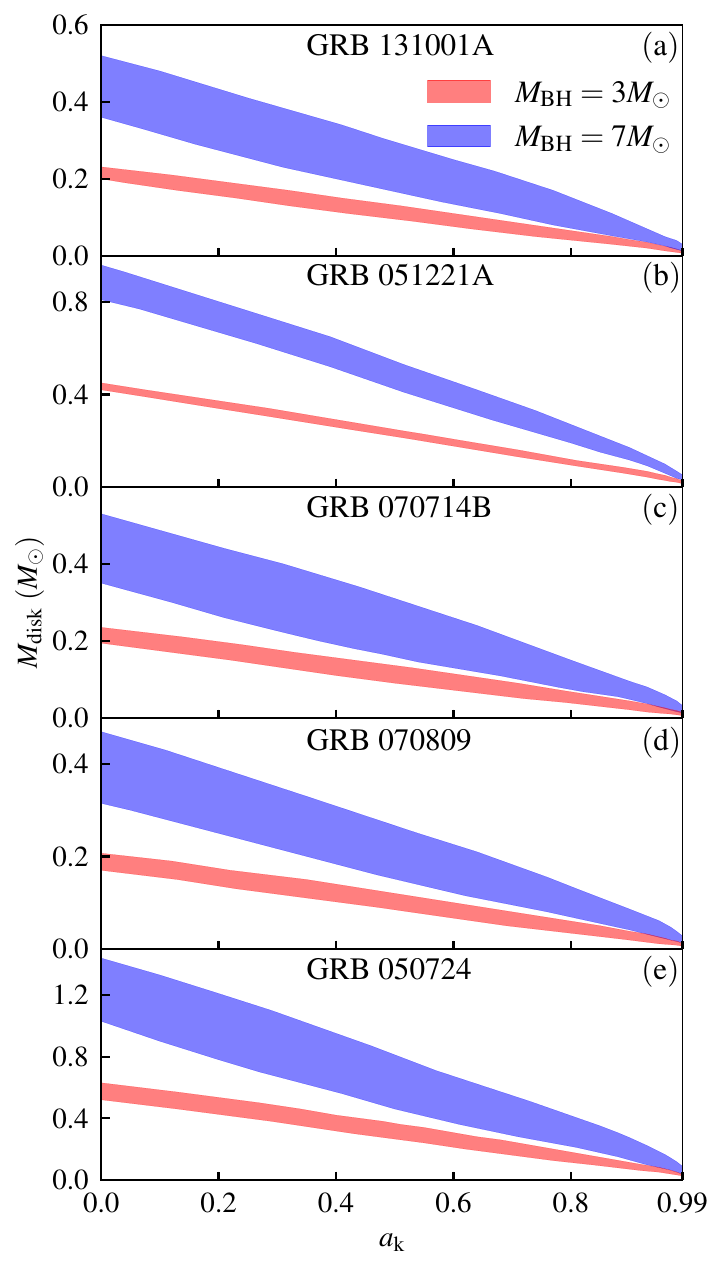}
    \caption{ Correlation between the black hole spin parameter ($a_{\rm k}$) and the disk mass ($M_{\rm disk}$) for different black hole masses ($M_{\rm BH}$). In each panel, the red and blue shaded regions represent the disk mass ranges corresponding to $M_{\rm BH}=3M_\odot$ and $M_{\rm BH}=7M_\odot$, respectively. The names of the SGRBs are indicated in each panel. See text for the details.
    }
    \label{fig3}
\end{figure}

In general, the central engine of GRBs is obscured by intense emission, making it challenging to observationally constrain its physical properties, namely black hole mass ($M_{\rm BH}$) and spin parameter ($a_{\rm k}$). As GRBs are often associated with compact object mergers, such progenitor systems impose constraints on the mass of the resulting black hole and the surrounding accretion disk \citep{Ruffert-Janka-1998, Ruffert-Janka-2001, Liu-etal-2015, Xie-etal-2016}. In the case of neutron star–neutron star (NS$+$NS) mergers, the mass of the remnant black hole is expected to be less than the total binary mass, typically $\lesssim 3 M_{\odot}$ \cite[]{Ruffert-Janka-1998, Ruffert-Janka-2001}. Conversely, in black hole–neutron star (BH$+$NS) mergers, the mass of the central BH remains of the order of a stellar mass scale \cite[$\sim 7M_\odot$,][]{Liu-etal-2015, Xie-etal-2016, Hayashi-etal-2021}. In both scenarios, numerical simulations suggest a typical disk mass of $M_{\rm disk}\sim\mathcal{O}(10^{-3}-10^{-1} M_{\odot})$ \cite[]{Kluzniak-Lee-1998, Ruffert-Janka-1998, Lee-Kluzniak-1999, Janka-etal-1999, Ruffert-Janka-2001, Shibata-etal-2003, Shibata-etal-2005, Shibata-Taniguchi-2006, Oechslin-Janka-2006, Kiuchi-etal-2009, Rezzolla-etal-2010, Hotokezaka-etal-2013, East-etal-2016, Most-etal-2019, Ruiz-etal-2020, Tootle-etal-2021, Hayashi-etal-2021,  Sun-etal-2022, Cokluk-etal-2024}. Indeed, the accretion rate onto the black hole is determined by both the disk mass and the active duration of the central engine. Upon considering the duration of burst ($T_{90}$) as the duration of activity of central engine, we estimate the mass accretion rate on to the black hole as \cite[]{Fan-Wei-2011},
 \begin{equation}
    \dot{m} \approx \frac{(1+z^{\prime})}{T_{90}} \left( \frac{M_{\rm disk}}{M_\odot}\right).
    \label{mdot}
\end{equation}

Using equations (\ref{dotE}) and (\ref{mdot}), one can estimate the energy deposition rate ($\dot{E}$) and mass accretion rate ($\dot{m}$) for a given accretion disk mass ($M_{\rm disk}$) associated with a specific GRB source. Adopting a representative value of $M_{\rm disk} = 0.2~M_{\odot}$, we compute $\dot{E}$ and $\dot{M}$ for three SGRBs for representation. For this analysis, the black hole mass is varied in the range $3 \leq M_{\rm BH}/M_{\odot} \leq 7$, and the spin parameter is taken within $0 \leq a_{\rm k}<1$. The viscosity parameter is kept fixed at $\alpha=0.01$ for illustrative purposes. We systematically explore different combinations of $M_{\rm BH}$ and $a_{\rm k}$ to obtain shock-induced global accretion solutions by suitably tuning the energy ($\mathcal{E}_{\rm in}$) and angular momentum ($\lambda_{\rm in}$), and compute the resulting neutrino annihilation luminosity $L_{\nu \bar{\nu}}$ (see equation \ref{Neutrino_anni_lum}), ensuring consistency with the observed values $L^{\rm obs}_{\nu\bar{\nu}}$ for each SGRB (see equation \ref{mean_out_pow}). The resulting parameter space in $a_{\rm k}-M_{\rm BH}$ plane is illustrated in Fig. \ref{fig2}, where the shaded regions enclosed by the dotted (grey), solid (red), and dashed (blue) curves correspond to GRB 131001A, GRB 051221A, and GRB 050724, respectively, with the corresponding accretion rates obtained from equation (\ref{mdot}) being ${\dot m}=0.223$, $0.221$, and $0.084$. In the figure, the shaded regions delineate the combinations of $a_{\rm k}$ and $M_{\rm BH}$ for which the assumed disk mass ($M_{\rm disk}$) can account for the observed luminosity of the SGRBs. Notably, for a given SGRB, the allowed range of $M_{\rm BH}$ systematically shifts toward higher values as the spin parameter $a_{\rm k}$ increases. This trend suggests that more massive black holes can satisfy the energy budget of SGRBs, provided the central engine possesses relatively higher spin.

Given that simulation studies suggest a broad range of possible disk masses \citep{Kluzniak-Lee-1998, Ruffert-Janka-1998, Lee-Kluzniak-1999, Ruffert-Janka-2001, Shibata-etal-2003, Shibata-etal-2005, Oechslin-Janka-2006, Kiuchi-etal-2009}, we therefore allow $M_{\rm disk}$ to vary freely and investigate its correlation with the spin parameter $a_{\rm k}$ while keeping the black hole mass fixed. This analysis is carried out for all five selected SGRBs such as GRB 131001A, GRB 051221A, GRB 070714B, GRB 070809 and GRB 050724, assuming two extreme central engine configurations, namely $M_{\rm BH} = 3M_{\odot}$ for NS$+$NS mergers and $M_{\rm BH} = 7M_{\odot}$ for NS$+$BH mergers. For various combinations of $M_{\rm disk}$ and $a_{\rm k}$, we tune the energy $\mathcal{E}_{\rm in}$ and angular momentum $\lambda_{\rm in}$ of the accretion flow to match the observed neutrino luminosity $L^{\rm obs}_{\nu\bar{\nu}}$. The allowed effective domains of the parameter spaces in $a_{\rm k}-M_{\rm disk}$ plane are depicted in Fig. \ref{fig3}, with red and blue shaded regions corresponding to $M_{\rm BH} = 3M_{\odot}$ and $7M_{\odot}$, respectively, across panels (a)–(e) for the SGRBs under considerations.

By analyzing the case of GRB 131001A, as shown in Fig. \ref{fig3}(a), we find a clear evidence of anti-correlation between $a_{\rm k}$ and $M_{\rm disk}$. This behavior arises because, when the disk mass is relatively low, the corresponding accretion rate decreases resulting in reduced density and temperature of the inflowing matter. Consequently, the accretion flow fails to attain the luminosity ($L_{\nu \bar{\nu}}$) required to match the observed value ($L^{\rm obs}_{\nu \bar{\nu}}$). To explain the observed luminosity under such conditions, the black hole must possess a higher spin, enabling matter to move deeper into the gravitational potential well and attain the necessary temperature and density. Furthermore, for a fixed value of $a_{\rm k}$, the admissible range of $M_{\rm disk}$ is considerably narrower for $M_{\rm BH} = 3~M_{\odot}$ than for $M_{\rm BH} = 7~M_{\odot}$. The reduced uncertainty in disk mass for $M_{\rm BH} = 3~M_{\odot}$ indicates that the present model framework seems to favor scenarios of NS$+$NS mergers over NS$+$BH merger. We next estimate the admissible range of disk mass required to power GRB 131001A for various spin parameters in the range $0 \le a_{\rm k} < 1$. For $M_{\rm BH} = 3 M_{\odot}$, the allowed disk mass lies within $0.007 \lesssim M_{\rm disk}/M_{\odot} \lesssim 0.231$, while for $M_{\rm BH} = 7 M_{\odot}$, it extends to $0.013 \lesssim M_{\rm disk}/M_{\odot} \lesssim 0.520$. For the remaining SGRBs under considerations, the ranges of $M_{\rm disk}$ for $M_{\rm BH} = 3 M_{\odot}$ and $7 M_{\odot}$ are listed in Table \ref{table-2}.

\begin{table}
\caption{Model estimates of the disk mass ($M_{\rm disk}$) corresponding to the lower and upper limits of the central engine mass ($M_{\rm BH}$) for the SGRB sources under consideration. Columns $1-4$ denote source name, observed neutrino annihilation luminosity ($L^{\rm obs}_{\nu \bar{\nu}}$), mass of the black hole ($M_{\rm BH}$) and disk mass ($M_{\rm disk}$). See text for the details.}
\begin{ruledtabular}
\begin{tabular}{lcccl}
Source &  $ L^{\rm obs}_{\nu \bar{\nu}}$ & $M_{\rm BH}$        &  $M_{\rm disk}$\\
       &  $\rm (erg~s^{-1})$  & ($M_{\odot}$) & ($M_{\odot}$)  \\
\hline
GRB 131001A  & $8.06\times10^{49}$   & 3 & $0.007 - 0.231$  \\
             &                       & 7 & $0.013 - 0.520$  \\
GRB 051221A  & $1.07\times10^{51}$   & 3 & $0.014 - 0.450$  \\
             &                       & 7 & $0.320 - 0.960$  \\
GRB 070714B  & $4.72\times10^{49}$   & 3 & $0.007 - 0.235$  \\
             &                       & 7 & $0.014 - 0.530$  \\
GRB 070809	 & $5.62\times10^{49}$   & 3 & $0.006 - 0.207$  \\
             &                       & 7 & $0.014 - 0.470$  \\
GRB 050724   & $9.50\times10^{49}$   & 3 & $0.026 - 0.620$  \\
             &                       & 7 & $0.037 - 1.440$  \\
\end{tabular}
\end{ruledtabular}
\label{table-2}
\end{table}

Generally, the remnant black hole formed after the merger accretes most of the binary mass, leaving a relatively low-mass disk \citep{Kiuchi-etal-2009, Cokluk-etal-2024}. However, for GRB 051221A and GRB 050724, the central black hole seems to host a comparatively massive disk ($\lesssim 1~M_{\odot}$) when $a_{\rm k} \rightarrow 0$ (see Fig. \ref{fig3}). The inferred disk masses in these cases exceed simulation predictions \citep[and references therein]{Kluzniak-Lee-1998, Ruffert-Janka-1998, Sun-etal-2022, Cokluk-etal-2024} likely because, in our model, the disk mass is treated as a free parameter tuned to reproduce the observed luminosity for all SGRB cases considered.

\section{\label{discussion_summery}Conclusions}

In this work, we investigate the properties of neutrino dominated accretion flows (NDAFs) to constrain the physical parameters of the central engine in five short GRBs. To achieve this, we solve the governing equations of NDAF focusing on the emergence of the shock phenomenon while incorporating neutrino emission as the dominant cooling process. In doing so, we adopt an effective potential that satisfactorily mimics the spacetime geometry around the central black hole  \cite[]{Dihingia-etal-2018}. We summarize the overall findings of the present work below.

\begin{itemize}
    \item For a given set of flow parameters, namely energy ($\mathcal{E}_{\rm in}$) and angular momentum ($\lambda_{\rm in}$), we obtain shock-induced global transonic accretion solutions across a broad range of black hole mass ($M_{\rm BH}$), spin parameter ($a_{\rm k}$), and accretion rate ($\dot{M}$). Using these shocked solutions, we compute the neutrino-antineutrino annihilation luminosity ($L_{\nu \bar{\nu}}$), which contributes to the energy powering GRBs (see Fig. \ref{fig1}).

    \item We compare the model estimated $L_{\nu \bar{\nu}}$ with the GRB power inferred from observation, thereby validating the capability of shock-induced NDAF models to account for the observed GRB energetics. Our analysis confirms that shock-induced accretion solutions are effective in explaining the broad range of GRB energies. Furthermore, by extending this framework, we attempt to constrain the key physical parameter, namely the black hole mass ($M_{\rm BH}$) for varied spin parameter ($a_{\rm k}$) of the GRB central engines (see Fig. \ref{fig2}), which otherwise remain challenging to determine directly from observations.

    \item By incorporating results from numerical simulations of NS$+$NS and NS$+$BH mergers, which is widely regarded as the progenitors of short GRBs, we employ the predicted range of post-merger disk masses ($M_{\rm disk}$) to constrain the spin ($a_{\rm k}$) of the central engine in GRBs. The analysis reveals that, for both merger scenarios, the spin parameter can be effectively bounded within the disk mass limits inferred from simulation results (see Fig. \ref{fig3}). This demonstrates that the present framework not only reproduces the observed energetics of short GRBs but also provides a physically motivated avenue for constraining the properties of their central engines. \\
    
\end{itemize}

Finally, we mention the limitations of this work, which is developed under several simplifying assumptions. To avoid the complexity of a full general relativistic treatment, we adopt an effective potential to approximate the spacetime geometry around the black hole. We consider an equation of state that accounts for gas and radiation pressure, while neglecting electron degeneracy and neutrino pressure. This simplification is reasonable because, for the chosen accretion rates (${\dot M} \le M_\odot~{\rm s}^{-1}$), electrons are not expected to become highly degenerate and neutrinos can escape the disk efficiently without significant trapping \cite[]{DiMatteo-etal-2002, Kohri-etal-2005, Kawanaka-Mineshige-2007, Chen-Beloborodov-2007, Liu-etal-2007, Xue-etal-2013, Xie-etal-2016}. Moreover, despite their ubiquity in accretion flows, magnetic fields are neglected in the present study. The inclusion of these effects lies beyond the scope of this work, however we plan to address these issues in future studies to be reported elsewhere.

\section*{Acknowledgments}

Authors acknowledge the support from the Department of Physics, IIT Guwahati, for providing the facilities to complete this work. 

\section*{Data Availability}

The data underlying this article will be available upon reasonable request.


\appendix

\section{Expression of coefficients in radial derivatives of $v$, $\lambda$ and $\Theta$}
 \begin{widetext}

 The detailed expression of $N$, $D$, $\lambda_{11}$, $\lambda_{12}$, $\Theta_{11}$ and $\Theta_{12}$ are expressed as, 
 \begin{align*}
    &\qquad \qquad \qquad N= \mathcal{A}_{x}\mathcal{B}_{\Theta}\mathcal{C}_{\lambda} - \mathcal{A}_{x}\mathcal{B}_{\lambda}\mathcal{C}_{\Theta} + \mathcal{A}_{\lambda}\mathcal{B}_{x}\mathcal{C}_{\Theta} - \mathcal{A}_{\Theta}\mathcal{B}_{x}\mathcal{C}_{\lambda} + \mathcal{A}_{\Theta}\mathcal{B}_{\lambda}\mathcal{C}_{x} - \mathcal{A}_{\lambda}\mathcal{B}_{\Theta}\mathcal{C}_{x},\\
    &\qquad \qquad \qquad D=
    \mathcal{A}_{v}\mathcal{B}_{\lambda}\mathcal{C}_{\Theta} - \mathcal{A}_{v}\mathcal{B}_{\Theta}\mathcal{C}_{\lambda} + \mathcal{A}_{\Theta}\mathcal{B}_{v}\mathcal{C}_{\lambda} - \mathcal{A}_{\lambda}\mathcal{B}_{v}\mathcal{C}_{\Theta} + \mathcal{A}_{\lambda}\mathcal{B}_{\Theta}\mathcal{C}_{v} -\mathcal{A}_{\Theta}\mathcal{B}_{\lambda}\mathcal{C}_{v},\\
    & \lambda_{11}=\frac{\mathcal{A}_{x}\mathcal{B}_{\Theta} - \mathcal{A}_{\Theta}\mathcal{B}_{x}}{\mathcal{A}_{\Theta}\mathcal{B}_{\lambda} - \mathcal{A}_{\lambda}\mathcal{B}_{\Theta}},\quad   \lambda_{12}=\frac{\mathcal{A}_{v}\mathcal{B}_{\Theta} - \mathcal{A}_{\Theta}\mathcal{B}_{v}}{\mathcal{A}_{\Theta}\mathcal{B}_{\lambda} - \mathcal{A}_{\lambda}\mathcal{B}_{\Theta}},\quad   \Theta_{11}=\frac{\mathcal{A}_{\lambda}\mathcal{B}_{x}-\mathcal{A}_{x}\mathcal{B}_{\lambda} }{\mathcal{A}_{\Theta}\mathcal{B}_{\lambda} - \mathcal{A}_{\lambda}\mathcal{B}_{\Theta}},\quad  \Theta_{12}=\frac{\mathcal{A}_{\lambda}\mathcal{B}_{v}-\mathcal{A}_{v}\mathcal{B}_{\lambda}}{\mathcal{A}_{\Theta}\mathcal{B}_{\lambda} - \mathcal{A}_{\lambda}\mathcal{B}_{\Theta}}.
\end{align*}
where
\begin{align*}
     \qquad   \mathcal{A}_\lambda &=1 + \lambda_{1} \Theta_{1}, &\mathcal{A}_\Theta &=\lambda_{1} \Theta_{2}, \quad 
                  &\mathcal{A}_{v}&= \frac{A_1S_1}{v~C_{s0}}, \quad &\mathcal{A}_{x}&=\lambda_{2}+\lambda_{1}\Theta_{3},\\
     \qquad  \mathcal{B}_{\lambda}&= R_{1} + R_{2} \Theta_{1},&\mathcal{B}_{\Theta}&= R_{2} \Theta_{2},
            \quad &\mathcal{B}_{v}&= R_{3} + R_{2} \Theta_{4},\quad &\mathcal{B}_{x}&= R_{4} + R_{2} \Theta_{3},\\
    \qquad  \mathcal{C}_{\lambda} &=  \mathcal{T}_1 -C_{s}^2  v \mathcal{T}_2, & \mathcal{C}_{\Theta}&=  \mathcal{T}_3 -C_{s}^2  v \mathcal{T}_4, 
            \quad &\mathcal{C}_{v}&=  \mathcal{T}_5 -C_{s}^2  v \mathcal{T}_6,  \quad & \mathcal{C}_{x}&= \mathcal{T}_7 -C_{s}^2  v \mathcal{T}_8
\end{align*}

\begin{align*}
    & \lambda_{1} =-\frac{2 C_{s} x\alpha}{v}, \quad \Theta_{1}=\frac{-C_s A_1 S_1\mathcal{F}_1 }{2 \mathcal{F}(2 C_{s}^2- A_1 S_1)}, \quad A_1 = \frac{11 \pi\bar{a} c^3 M_{BH}^2 m_{\rm e}^2 }{3 k_{\rm B} \dot{M}} , \quad S_1= v H\Theta^4 \sqrt{\Delta}, \quad \mathcal{F}_{1}= g_1 \frac{(\lambda\ \Omega_\lambda+\Omega)}{(1-\lambda \Omega)^2},\\
    & g_1=\frac{(x^2+a^2_{\rm{k}})^2+2 \Delta a^2_{\rm{k}}}{(x^2+a^2_{\rm{k}})^2-2 \Delta a^2_{\rm{k}}}, \quad 
    \Omega_\lambda =\frac{\partial \Omega}{\partial \lambda}, \quad C_{\rm s}= \frac{Y_{1}+ \sqrt{Y_{1}^2 + 4 Y_{2}} }{2}, \quad Y_{1}= A_1 v  \sqrt{\frac{ \Delta x^3}{\mathcal{F}}} \Theta^4, \quad Y_{2}= \frac{m_{\rm e}}{m_{\rm p }}\Theta, \\
    & \lambda_{2}=\frac{\alpha(C_{\rm s}^2+v^2)}{v}\left(\frac{x}{2}\frac{\Delta^{\prime}}{ \Delta}-2\right), \quad \Delta^{\prime}=\frac{d\Delta}{dx},  \quad \Theta_{3}=\frac{A_1 S_1}{ C_{s0}}\left(\frac{3}{2 x} - \frac{\mathcal{F}_{2}}{2\mathcal{F}} \right) + \frac{S_1 A_1 \Delta^{\prime}}{ 2 C_{s0} \sqrt{\Delta} }, \quad C_{s0}= 2 C_s -\frac{A_1 S_1}{C_s}, \\
    & \mathcal{F}_{2}= \frac{1}{1-\lambda\Omega} \left(\frac{d g_1}{d x} + \gamma_{\phi}^2 \lambda g_1 \frac{d\Omega}{dx}\right) , \quad R_{1}= C_{\rm s}^2 \frac{\mathcal{F}_{1}}{2\mathcal{F}}, \quad R_{2}= C_{\rm s},\quad R_3= v-\frac{C_{\rm s}^2}{v}, \quad \Theta_{4}= \frac{A_1 S_1 }{v C_{s0}},\\
    & R_{4}= \frac{d \Phi^{ \rm{eff}}}{dx} +C^2_{\rm s}\left(\frac{\mathcal{F}_2}{2 \mathcal{F}}- \frac{\Delta^{\prime}}{2 \Delta} -\frac{3}{2x}\right), \quad \mathcal{T}_1= \Gamma_1  -3A_1S_1 v \left( \frac{\mathcal{F}_1}{2 \mathcal{F}}  +\frac{\Theta_{1}}{C_s} \right), \quad \Gamma_1= -\alpha(v^2 + C_{s}^2)x\Omega_\lambda ,\\
    & \mathcal{T}_2= \frac{\mathcal{F}_{1}}{2\mathcal{F}} -\frac{\Theta_{1} }{C_s} , \quad \mathcal{T}_3  =  \frac{ v}{\gamma-1} \frac{m_{\rm e}}{m_{\rm p}} + 3A_1 S_1 v \left(\frac{4}{\Theta} +\frac{\Theta_2}{C_s}\right), \quad  \mathcal{T}_4=  -\frac{\Theta_{2}}{C_{s}}, \quad \mathcal{T}_5 =  3 A_1 S_1   +\frac{3 A_1 S_1 \Theta_{4} v}{C_s}, \quad \mathcal{T}_6 =  -\frac{\Theta_4}{C_s} -\frac{1}{v},\\
    & \mathcal{T}_7 =  \Gamma_2- \frac{Q_\nu}{\Sigma} + \mathcal{C}_{1},~\Gamma_2= -\alpha(v^2 + C_{s}^2)x \frac{d\Omega}{dx}, \quad \mathcal{C}_{1}= \frac{3 A_2 S_1 v}{2} \left(\frac{\Delta^{\prime}}{\Delta}- \mathcal{F} \mathcal{F}_2 + \frac{2 \Theta_3}{C_s} +\frac{3}{x} \right),\quad \mathcal{T}_8 = -\frac{\Theta_3}{C_s} +\frac{\mathcal{F}_2}{2 \mathcal{F}}- \frac{\Delta^{\prime}}{2 \Delta} -\frac{3}{2x}.
\end{align*}

\end{widetext}


%

\end{document}